\documentclass[11pt]{article}
\usepackage[english]{babel}
\usepackage[export]{adjustbox}

\usepackage[letterpaper,top=2cm,bottom=2cm,left=3cm,right=3cm,marginparwidth=1.75cm]{geometry}

\usepackage{amsmath}
\usepackage{graphicx}
\usepackage[colorlinks=true, allcolors=blue]{hyperref}
\usepackage{booktabs}
\usepackage{caption}
\usepackage{subcaption}
\usepackage{colortbl}

\usepackage{authblk}

\title{\textbf{Drivers of the decrease of patent similarities from 1976 to 2021}}
\author[1]{Edoardo Filippi-Mazzola\footnote{edoardo.filippi-mazzola@usi.ch}}
\author[1]{Federica Bianchi}
\author[1]{Ernst C. Wit}
\affil[1]{Institute of Computing, Università della Svizzera italiana, Lugano, Switzerland}
\date{}                     

\begin{document}
	\maketitle

\begin{abstract}
The citation network of patents citing prior art arises from the legal obligation of patent applicants to properly disclose their invention. One way to study the relationship between current patents and their antecedents is by analyzing the similarity between the textual elements of patents. Many patent similarity indicators have shown a constant decrease since the mid-70s. Although several explanations have been proposed, more comprehensive analyses of this phenomenon have been rare.

In this paper, we use a computationally efficient measure of patent similarity scores that leverages state-of-the-art Natural Language Processing tools, to investigate potential drivers of this apparent similarity decrease. This is achieved by modeling patent similarity scores by means of generalized additive models. We found that non-linear modeling specifications are able to distinguish between distinct, temporally varying drivers of the patent similarity levels that explain more variation in the data ($R^2\sim 18\%$) compared to previous methods. Moreover, the model reveals an underlying trend in similarity scores that is fundamentally different from the one presented previously. 

\end{abstract}

\section{Introduction}

Understanding the characteristics of ground-breaking innovations is crucial for technology-based firms striving for success~\cite{henderson1990architectural}. Patent indicators serve this purpose and support, among the others, the development of product strategies~\cite{wang2019novelty, park2013identification}, monitoring of existing technological trends \cite{wang2010identifying}, the detection of promising opportunity of investments~\cite{yoon2012detecting}, the assessment of technological impact of novel applications~\cite{verhoeven2016measuring, veugelers2019scientific}, and recognizing similar technologies~\cite{an2021improved, kuhn_patent_2020, whalen_patent_2020, an2018deriving}.

Patent indicators using institutional classifications and citation information are predominant~\cite{gress_properties_2010, verhoeven2016measuring, acemoglu_innovation_2016, veugelers2019scientific} in patent analysis. Patent classification systems like the International Patent Classification (IPC) are usually processed for identifying patents that are technologically similar. However, technological relatedness may not be fully captured by sharing the same patent class. Despite numerous methods for analyzing technological relatedness and closeness based on such classes~\cite{yan_measuring_2017}, their usage can be problematic when patents need to be identified, compared, or matched with similar technologies.

In contrast, patent indicators using patent descriptions and lexical contents are less common in patent analysis. A keyword-based approach using frequency and co-occurrence of contents is typically used for computing the technological similarity between pairs of patents~\cite{younge_patent--patent_2015}. Within the set of patent indicators, patent similarity is a fundamental measure in the evaluation of technological novelty~\cite{wang2019novelty} and infringement risks associated with others using or selling inventions without authorization~\cite{an2021improved}. Patent descriptions can be mined for combinations of words and unique expressions for text-based indicators for patent similarity. This transforms unstructured textual data into actionable knowledge through latent relationships between patent documents~\cite{immordino2019comparing}.

With the development of new and more sophisticated deep learning techniques, Natural Language Processing (NLP) tools have been proven to provide valid alternatives to canonical technology class measurements. The idea is to use the textual elements of patents as inputs for defining vectors of similarity. In this way, it is possible to use continuous distance measures between any two patents, e.g., Euclidean distance, cosine similarity, or Mahalanobis distance to measure patent (dis)similarity. Although the idea of mapping patents into a vector space can be traced back to~\cite{jaffe_technological_1986, jaffe_characterizing_1989}, only recently these methods have been applied to patent analysis. For example, \cite{younge_patent--patent_2015} used a bag of words methodology~\cite{turney_frequency_2010} to develop a machine-automated patent-to-patent similarity measure based on the technical descriptions of patent applications. Adopting the same approach, ~\cite{kuhn_patent_2020} analyzed pairs of patent citations in the US between 1975 and 2014. Simple vocabulary-based approaches of textual similarity scores across citing and cited patents, may contain major drawbacks caused by the sparsity of the output matrix. Although there have been developments to address this weakness~\cite{deerwester_indexing_1990}, a neural network (NN) approach, such as the one proposed by~\cite{whalen_patent_2020}, is preferred as semantics and context are prioritized within the estimated positional embeddings. The introduction of language based NN models has opened up the way for more complex applications within patent similarity analyses. While early contributions have focused on patent abstract data for correctly classifying patents into their technological classes~\cite{Lee_classification_2020,bekamiri_sbert_classification_2021}, the focus is now shifting towards mapping patents into multidimensional spaces to detect patterns and gain relational insights. In this regard, \cite{hain_text-embedding-based_2022} proposed using a K-nearest-neighbors algorithm to spot closely related patents by training a Word2Vec NN model~\cite{word2vec_2013} on 48 million abstracts. Regardless of the amount of data processed, the computational cost of these approaches are high. Instead, the current availability of generic models pre-trained on massive corpora is rapidly increasing \cite{Liu_Cai_Guo_Chen_2021}. This has enabled researchers to unlock vast complex natural language models with fewer computational resources, paving the way for a new set of tools. 

In the context of textual similarity analysis in patent citations,~\cite{kuhn_patent_2020} and~\cite{whalen_patent_2020}, noted a decrease in the average textual similarity per year between citing and cited patents. The aim of this manuscript is to investigate the drivers of patent similarity decline during a period of approximately forty years, from 1976 to 2021, with 1976 the year when the \emph{US Patent Trading Office} (USPTO) started collecting the full text for all granted patents in digital databases. Previous studies of the decrease of patent similarity attribute this drop to fundamental changes that occurred in the data generation process. \cite{kuhn_information_2010} claim that legal changes in the applicant's duty of disclosure has led to a drastic increase in the number and scope of cited references. As a consequence, more citations have been included that are further afield from the citing patent. Pursuing this hypothesis, \cite{kuhn_patent_2020} show how the skewed distribution of backward citations has become less informative for research practices, as a small minority of patent applications are now generating a large majority of patent citations in the overall citation network.

We propose to use pre-trained models to compute the embeddings. In this sense, we avoid any computational procedure by proposing instead a ready-to-use approach for computing similarity scores. We focus on patent abstracts that contain the most concise information regarding the patenting technology \cite{choi_deep_2022,hain_text-embedding-based_2022}. Thanks to the reduced size of the abstract corpus, we are able to compute the positional embeddings via a pre-trained SBERT model in a reasonable amount of time. We encode the entire set of roughly 10 million abstracts into fixed sized vectors and compute the vector of similarity scores across 100 million patents citations through a parallelized lazy loading scheme. 

We will first describe the USPTO patent data on which we base our analysis. We then describe the SBERT embedding of the abstracts data and the calculation of the patent similarity scores. The scores confirm the downward trend in the patent similarity scores. Then we propose various generalized additive models with the aim of detecting the drivers of patent similarity over time, in particular, whether this is a temporal endogenous process or due to exogenous patent attributes. In contrast to previous studies, our approach also aims to resolve the problem of the temporal boundary of the citation network by considering the time lag between the citing and the cited patents. 

\section*{Materials and methods}

\subsection*{USPTO patent data}

Intellectual property history can be traced back to the 19th century when the first patenting office was established in Paris. Since then the patenting documentation has evolved and the availability of patent data has grown dramatically. One of the main challenges of analyzing patent data is retrieving the required information from the large amount available. Moreover, patents are legal documents, mostly consisting of textual elements. Unfortunately, the non-availability of standardized patent formats through the years has caused difficulties in building standardized data bases. Moreover, the juridical procedures of patenting are country-specific. This creates inconsistencies in the data from different countries, as some patenting offices will use different citation procedures. A striking example is a distinction in the citation process between the USPTO and the European Patent Office (EPO). Both the USPTO and the EPO require applicants to fulfill their duty of disclosure by citing all the required prior arts. The examiner committee of the USPTO adds citations to the application by integrating all those prior arts that are considered relevant for the patent to be correctly disclosed. On the other side, the EPO examiner committee does not include any further citations in the examination process. The committee limits its range of action by evaluating the validity of the patent combined with the disclosed prior arts. From this perspective, a combined analysis of multiple patenting offices' data would result in unreliable conclusions. 

For this reason, we focus our analysis exclusively on patents that have been issued by the USPTO from January 1976 up to September 2021. Starting from 1976, the USPTO has created an online public repository storing all the issued patents, including guidelines for data quality and standardization in the textual component of submitted legal documents. Although the USPTO data are broadly available across different periods, we have noted that most common repositories contain many inaccuracies. Such issues are usually the result of heavy preprocessing procedures used to combine, correct, or fill missing values from distinct sources to integrate the range of data that the USPTO provides publicly. To retain the highest quality possible in our dataset, we avoid third-party preprocessing and download data directly from the USPTO digital repository (\url{https://bulkdata.uspto.gov/}). After downloading the required XML files, these were processed and combined to obtain CSV files through an open-source software tool (available at: \url{https://github.com/iamlemec/fastpat}).

Our dataset consists of a time-stamped citation network along with patent attributes. For each granted patent we consider its backward citations, and for each patent in the dataset we include International Patent Classification (IPC) codes. In line with the network science vocabulary, we refer to citing patents as senders and to cited patents as receivers.

\subsection*{The unreliability of institutional classification schemes}

Patent classification schemes like those illustrated in the IPC Table \ref{tab:ipc} are designed for examiners to ease the examination process of patent applications by rapidly searching for similar or related technologies. Studies on innovation use such instruments to analyze potential technological patterns, usually through similarity levels derived from co-class proximity measures~\cite{yan_measuring_2017}.

\begin{table}[!ht]
		\centering
		\caption{International Patent Classification (IPC) scheme for a generic patent classified as \emph{A01C 3/04}.}
		\label{tab:ipc}
		\begin{tabular}{|ll|l|l|l|} 
			\hline
			\multicolumn{1}{|l|}{\textbf{A}}       &                      &                      &                      &                              \\
			\multicolumn{1}{|l|}{\textit{Section}} & \textbf{A01}         &                      &                      &                              \\ 
			\cline{1-1}
			& \textit{Class}       & \textbf{A01C}        &                      &                              \\ 
			\cline{1-2}
			& \multicolumn{1}{l}{} & \textit{Subclass}    & \textbf{A01C 3/00}   &                              \\ 
			\cline{1-3}
			& \multicolumn{1}{l}{} & \multicolumn{1}{l}{} & \textit{Group}       & \textbf{\textit{A01C 3/04}}  \\ 
			\cline{1-4}
			& \multicolumn{1}{l}{} & \multicolumn{1}{l}{} & \multicolumn{1}{l}{} & \textit{Subgroup}            \\
			\hline
		\end{tabular}

	\end{table}

It has been argued that institutional classification schemes do not offer a reliable picture of patent similarity.~\cite{younge_patent--patent_2015} explain how many sources of bias may emerge when comparing patents through the technological classes they belong to. On the one hand, patent classes are not fixed -- i.e., new technological classes may be created and old classes may be merged, split, and/or reassigned in a way that affect the depth of technological spaces. On the other hand, the classes may be too broad or too tight, leading to inaccurate comparisons. 

\begin{figure}[!h]
	\includegraphics[width=\textwidth]{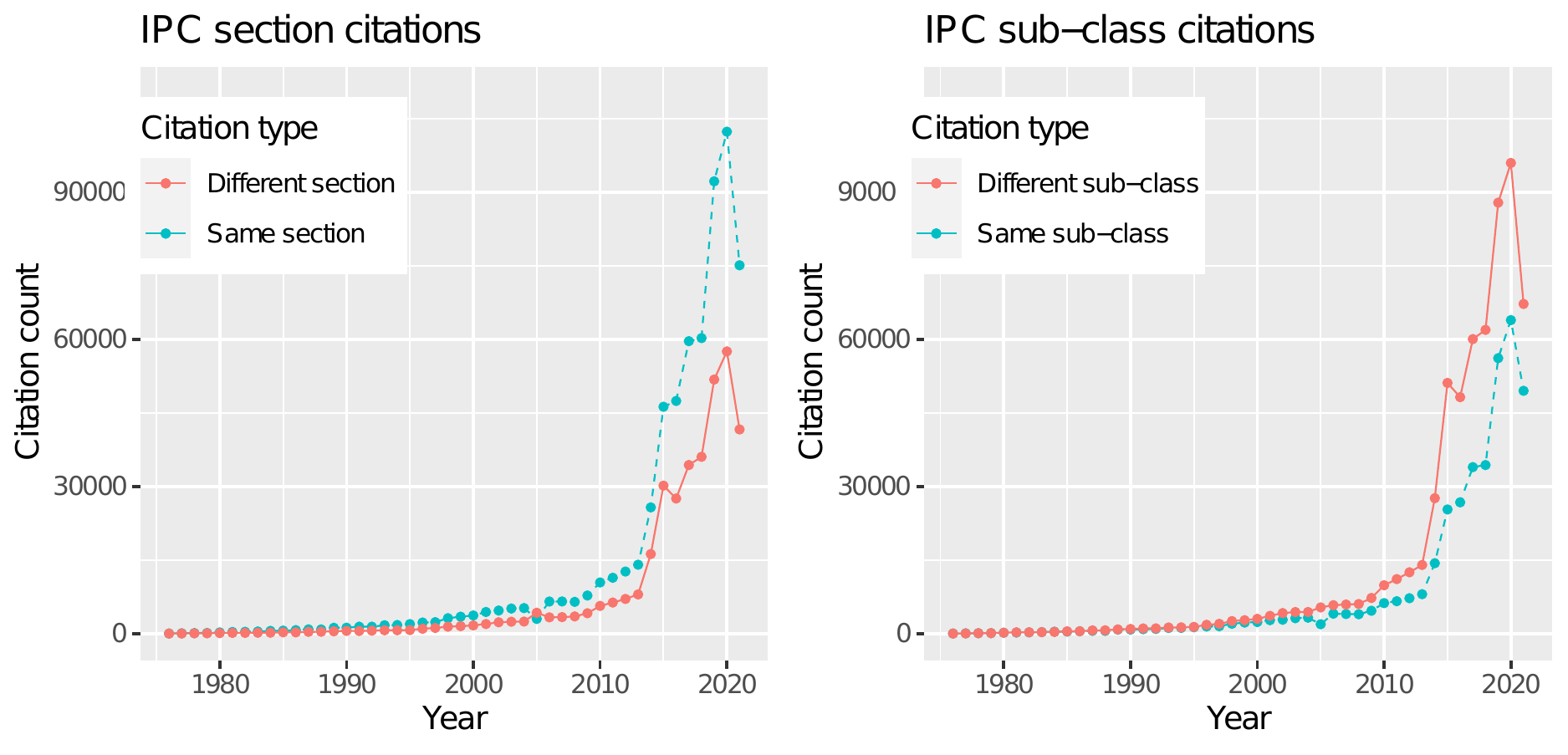}
	\caption{{\bf IPC citations comparison.}
		Amount of citations within the same class (left) and sub-class (right) during the observation period.}
	\label{fig1}
\end{figure}

We compare a random sample of 1 million citations through their \emph{sections} and \emph{sub-classes} as defined in the IPC classification, where sections are the broader category and sub-classes are the preferred level of analysis in empirical applications. Fig~\ref{fig1} clearly shows that any measure of patent similarity based on institutional classifications suffers from a selection bias in the hierarchy of classification layers. In fact, while technology classes tend to self-cite, which produce higher similarity scores, technology sub-classes tend to cite outside of their area.

\subsection*{Patent similarity based on pre-trained SBERT}

NLP tools can be used to interpret textual and translate it into a mathematical form in order for other algorithms to accomplish predefined tasks. What has been a revolution for NLP was the introduction of the Transformer architecture~\cite{vaswani_attention_2017} in a field that was previously dominated by Recurrent Neural Networks and Long-Short-Term-Memory networks. The great step that made the Transformers the new go-to tools for NLP is the focus on attention mechanisms~\cite{bahdanau2014attention} which replaced recurrence functions with a large number of parameters. Instead of processing the sentence sequentially (or word-by-word), the attention mechanism processes the entire sequence to give weights to the input. In this sense, it decides how much each word in the input is associated to the sentence. In this way, it runs a probabilistic-like approach that prioritizes certain parts over others. Transformers then combine an encoder-decoder architecture which solely relies on the attention mechanism to forward more parts of the input sequence at once (see~\cite{rothman2021transformers} for an overview).

The \emph{Bidirectional Encoder Representations from Transformer} (BERT)~\cite{devlin_bert_2019} takes this concept and extends it by using the context coming from both sides of the current analyzed part of the input. This change is significant as often a word may change meaning while the sentence develops. Each word added augments the overall meaning of the word being analyzed. The more words that are present in total in each input sequence, the more ambiguous the word in focus becomes. BERT accounts for the augmented meaning by reading the input bidirectionally, accounting for the effect of all other words in the input on the focus word and eliminating the left-to-right shift that biases words towards a certain meaning as the sentence progresses.

Although BERT outperforms any other benchmark that was set by previous NLP tools to encode the meanings of words into queries, it does not perform well when it comes to comparing similarities of entire sentences. A large disadvantage of the BERT network structure as presented by~\cite{devlin_bert_2019} is that independent sentence embeddings are not computed, which makes it difficult to derive sentence embeddings from BERT. In response, ~\cite{reimers-2019-sentence-bert} modified the standard BERT architecture for semantic textual similarity, called Sentence-BERT (or SBERT), while also reducing computing time. The main difference with the regular BERT architecture is encoding the semantic meaning of whole sentences instead of individual words. The SBERT architecture is characterized by a so-called twin network, which allows it to process two sentences simultaneously. The twins are identical down to every parameter, which allows it to think of this architecture as a single model used multiple times. At the end of the SBERT pipeline, the model contains a final pooling layer that enables the creation of a fixed-size representation for input sentences of varying lengths. With this, it is possible to encode documents into fixed-sized vectors, while taking their semantics into account. 

The downside of models based on Transformer architectures is that these are among the most computationally intensive Neural Networks to train. The peer-to-peer \emph{Hugging Face} repository solve this deficiency and allows researchers to upload trained models in an open-source fashion. With this tool, access to deep and complex neural networks are in within the reach of every user. Moreover, pre-trained SBERTs have two other advantages over competing models. The first is that SBERT pre-trained models uploaded on \emph{Hugging Face} are trained either for specific or general purposes. General purpose models are trained on billions of generic documents, which grants flexibility to use SBERT for any task. The second one is the ease of use granted by the package \emph{Sentence Transformers}, which simplifies the procedure of creating and downloading the pre-trained weights from \emph{Hugging Face} with a few lines of code. These two reasons, combined with the established benchmarks that SBERT has set in the field of NLP, make this the go-to model for our task. The \emph{Sentence Transformers} package in python gives access to pre-trained models from the \emph{Hugging Face} repository that encodes sequences into fixed-sized vectors. From this library, we downloaded a model, trained and fine-tuned on more than one billion public documents that encodes texts into vectors of size 384. 

\begin{figure}[!h]
	\includegraphics[width=\textwidth]{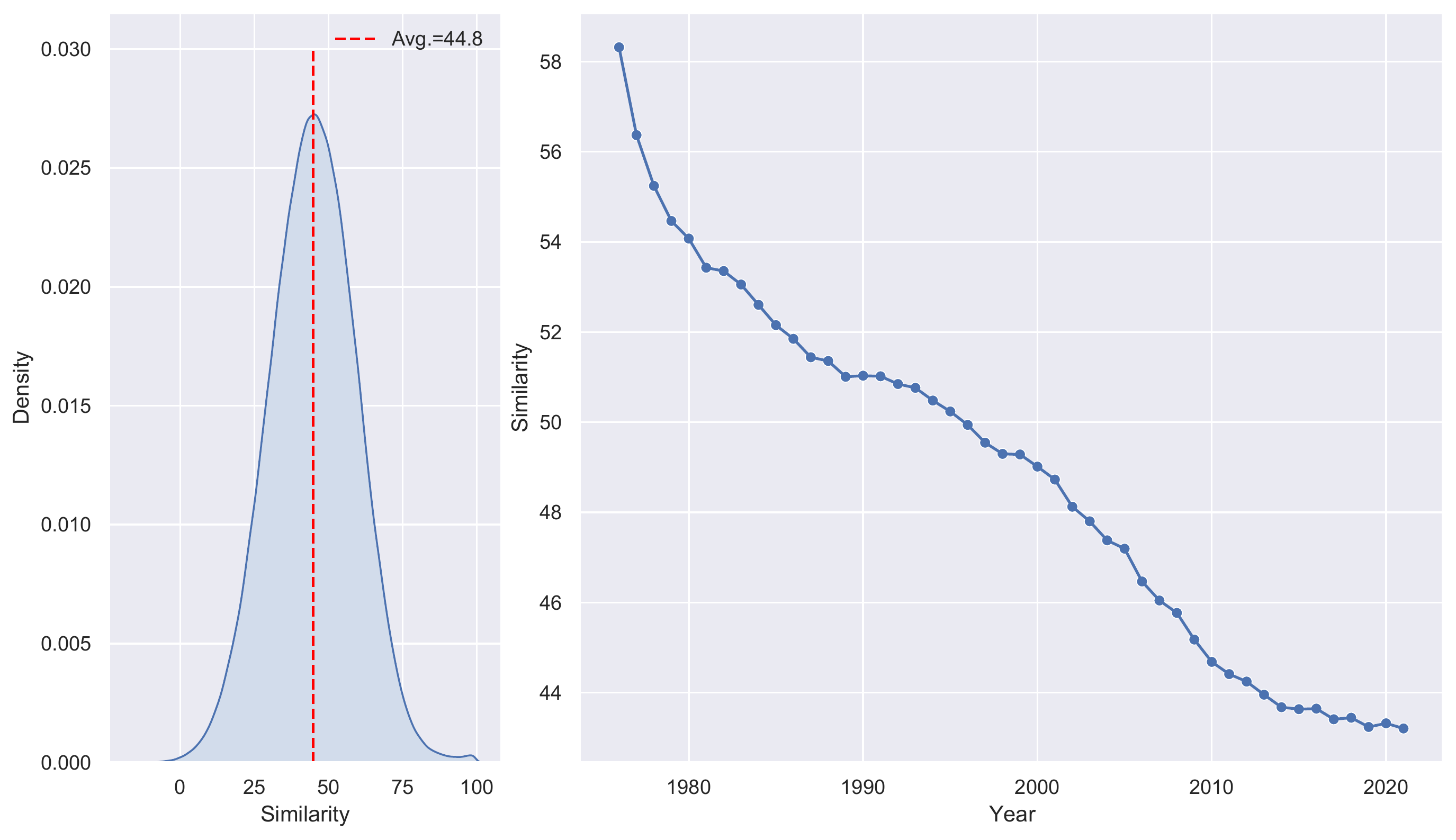}
	\caption{{\bf Citation similarity.} 
		\emph{Left}: similarity distribution. \emph{Right}: average citation similarity per year.}
	\label{fig2}
\end{figure}

Similarly to~\cite{whalen_patent_2020} and~\cite{choi_deep_2022}, we removed non-utility patents (such as plants or designs) from our data when computing embeddings. In this way, we encoded approximately 7.5 million patents into a fixed-size space through SBERT. By parallelizing a scheme of lazy loading procedures, we managed to compute the patent similarity scores for almost 100 million patent citations within minutes. Confirming results from previous studies. Fig~\ref{fig2} shows that the average similarity per year between citing papers is decreasing over time. The cosine similarity scores range between -1 and 1, we multiplied them by 100 for ease of representation. 

Embeddings were computed on a cluster node on parallel with two NVIDIA graphic processor units, models GeForce GTX 1080 Ti with 3584 CUDA cores and 100 GB of RAM. Computation time stands within one hour (more information on the resources that were used can be found on \url{https://intranet.ics.usi.ch/HPC}). Thanks to the usage of a pre-trained architecture, there is no training involved in the computation of the embeddings. This leaves the machine resources to take advantage of this method by taking the abstracts as input and provide only one forward passage among the neural network. Given this simplified procedure, the same results can be obtained within a reasonable time using less powerful computational resources -- e.g., standard Colab notebooks.

\subsection*{Modeling similarity scores through GAMs}

It has been claimed that due to the changes in the generative process of patent citations, these citations have become less informative and representative~\cite{kuhn_patent_2020}. We argue instead that with the correct application of informative statistical models, it is still possible to gain important insights on the main drivers of the decrease of patent similarity. As such, we argue that the backward citation process still plays a major role in determining the technological proximity of patents and the direction in which the network of citations is expanding. 

We propose to model textual similarity scores through GAMs by extending the approach of~\cite{kuhn_patent_2020}. GAMs can be used to estimate non-linear effects of covariates on the dependent variable. More in detail, while in linear models the predictor is a weighted sum of the $p$ covariates, $\sum_{j=1}^p \beta_j x_j$, in GAMs this term is replaced by a sum of functions, e.g. $\sum_{j=1}^p \sum_{l=1}^q \beta_j b_l(x_j)$, where the $b_1(.),\dots, b_q(.)$ are specific parametric basis functions -- e.g. smoothing splines or complex polynomial splines. Essentially, GAMs are particularly useful for uncovering nonlinear drivers of some process~\cite{hastie_gams_1986}.

\paragraph{Model 0.} Using this modeling technique, the decrease of patent similarity can be visualized by a simple GAM with the patent publication date as unique covariate modeled by a smooth term \emph{(Model 0)}.

\begin{figure}[!h]
	\includegraphics[width=\textwidth]{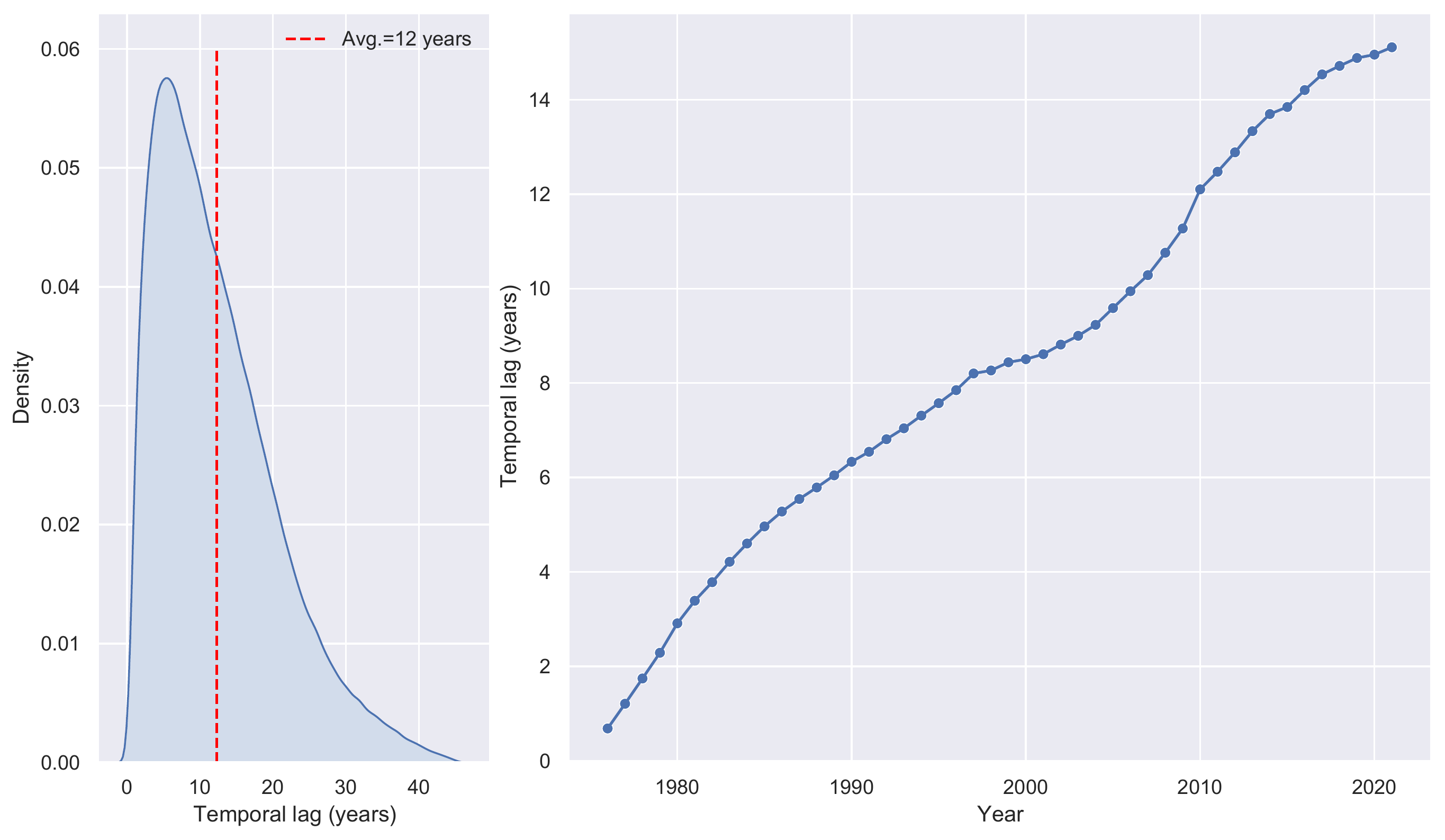}
	\caption{{\bf Temporal lag.} 
		\emph{Left}: temporal lag distribution. \emph{Right}: average temporal lag by year.}
	\label{fig3}
\end{figure}

\paragraph{Model 1.}The average decrease of similarity in backward citations is associated with an increase in the average temporal lag elapsed between citing and cited patents (see Fig~\ref{fig3}). This result seems to suggest that applicants and examiners cite prior arts which are increasingly temporally distant to their application/grant time. By itself, this effect could be a source of the reduction of similarity levels as the innovation process gives reasons to believe that temporally distant technologies are less similar to newer ones. In addition, in a period of approximately forty years, the legal and technical language has seen some important changes. Although the usage of SBERT should eventually mitigate the change of language as the model would account for context and semantics, the language evolution follows the technological development present inside patents, thus increasing the reduction of a potential similarity effect.

In our model specification, the temporal component of patent citations is captured by the covariate \emph{temporal difference}. Its effect on patent similarity is modeled through a smoothing spline of the time lag (in days) between issued dates of the cited and citing patent. Together with the sender publication date, we account for temporal effects that address the impact of two fundamental temporal components that influence the similarity levels \emph{(Model 1)}. 

\paragraph{Model 2.}~\cite{kuhn_information_2010} argued how legal changes that occurred within the early 2000s amplified the incentives to disclose, increasing the number of citations per issued patent. This idea is also discussed in~\cite{kuhn_patent_2020}, where a negative effect for the number of backward citations has been observed. However, the inflation of the number of citations during the last period may well be a bias in which a linear effect does not properly take this into account. We address this issue by adding to the model a further covariate fitted through a smooth term: the \emph{backward citation count}. With this explanatory variable, we correct for the increasing number of backward citations done by a given citing party. Furthermore, we consider the type of applicant who is providing the citation, discriminating between organizations (both profit and non-profit) and privates owners. The reason for this is straightforward: if there is an inflation in the number of citations, these could be more present within organizations than by private applicants as the former tend to cite more on average. We added three distinct fixed effects in the form of binary variables: \emph{is the same organization} if the citing and the cited company of the patent coincide, \emph{is citing party an organization} and \emph{is cited party an organization} if either the owner of the cited is an organization \emph{(Model 2)}.

\paragraph{Model 3.}To complete our analysis, we introduced effects related to the IPC for both the citing and the cited party. As we explained, the usage of technological classes for the assessment of patent similarity is disputable~\cite{younge_patent--patent_2015}, but still very important as a common source of knowledge for applicants and examiners. Following~\cite{yan_measuring_2017}, for each pair of citing/cited patents we computed the \emph{Jaccard index} of individual components in the IPC scheme. What we obtained are five distinct distributions that summarize the technological relatedness between two patents at different levels of the hierarchical classification system.

\section*{Results}

\begin{figure}[!h]
	\includegraphics[width=\textwidth]{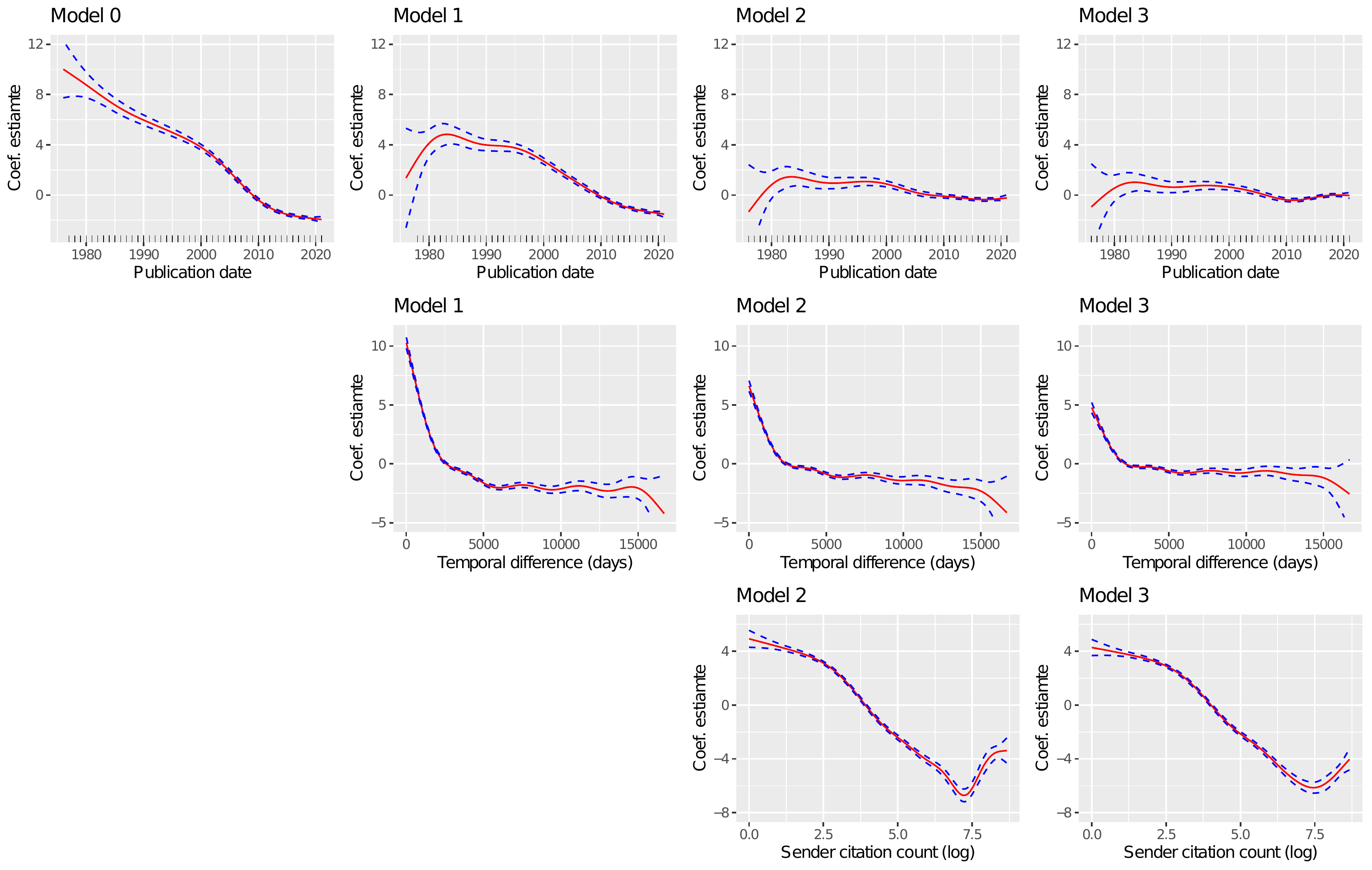}
	\caption{{\bf Models splines.} 
		Smoothing splines estimates for the three fitted GAMs. \emph{Columns}: Effects fitted per model. \emph{Rows}: Comparison of the same effects for the distinct models. Assessment criterions can be found at the bottom of Table \ref{tab:results}.}
	\label{fig4}
\end{figure}

The naive \emph{Model 0} is in line with previous studies, suggesting that patent similarity is decreasing with time (Fig~\ref{fig4}). However, this view fails to take into account various confounding effects. In turn, we will focus our attention to the time lag between the citing patent and those patents it cites (\emph{Model 1}), exogenous information associated with citing and cited patent owners, the increasing number of citations per patent (\emph{Model 2}), and finally the IPC classification of the patents (\emph{Model 3}). Empirical results are reported in Table \ref{tab:results}. For each model specification, Fig~\ref{fig4} compares the estimated splines associated with each effect.

\begin{table}[!h]
	\centering
	\caption{Coefficient estimates for the fixed parametric effects in the three model specifications. Refer to Fig~\ref{fig4} for the smoothing splines terms. Asterisk symbol representes the levels of statistical significance, while values in parenthesis are the related standard errors. Model assestment criterions can be found at the bottom of the table. For our purpouses, we compared three different criterions: Akaike Information Criterion (AIC), Generalized Cross Validation (GCV) criterion and the Deviance explained (R-squared).}
	\label{tab:results}
	\begin{tabular}{|l|l|l|l|l|} 
		\hline
		\rowcolor[rgb]{0.749,0.749,0.749}                             & \textbf{Model 0}                                                 & \textbf{Model 1}                                                  & \textbf{Model 2}                                                   & \textbf{Model 3}                                                    \\ 
		\hline
		&                                                                  &                                                                   &                                                                    &                                                                     \\ 
		\hline
		Intercept                                                     & \begin{tabular}[c]{@{}l@{}}$45.16^{***}$\\$(0.149)$\end{tabular} & \begin{tabular}[c]{@{}l@{}}$45.16^{***}$\\ $(0.015)$\end{tabular} & \begin{tabular}[c]{@{}l@{}}$45.933^{***}$\\ $(0.062)$\end{tabular} & \begin{tabular}[c]{@{}l@{}}$41.601^{***}$\\ $(0.065)$\end{tabular}  \\ 
		\hline
		Is the same organization?                                          &                                                                  &                                                                   & \begin{tabular}[c]{@{}l@{}}$8.758^{***}$\\ $(0.055)$\end{tabular}  & \begin{tabular}[c]{@{}l@{}}$7.883^{***}$\\ $(0.053)$\end{tabular}   \\ 
		\hline
		Is sender an organization?                                          &                                                                  &                                                                   & \begin{tabular}[c]{@{}l@{}}$-1.225^{***}$\\ $(0.056)$\end{tabular} & \begin{tabular}[c]{@{}l@{}}$-1.068^{***}$\\ $(0.053)$\end{tabular}  \\ 
		\hline
		Is receiver an organization?                                        &                                                                  &                                                                   & \begin{tabular}[c]{@{}l@{}}$-1.509^{***}$\\ $(0.045)$\end{tabular} & \begin{tabular}[c]{@{}l@{}}$-1.226^{***}$\\ $(0.043)$\end{tabular}  \\ 
		\hline
		Jaccard index: section                                        &                                                                  &                                                                   &                                                                    & \begin{tabular}[c]{@{}l@{}}$2.248^{***}$\\ $(0.056)$\end{tabular}   \\ 
		\hline
		Jaccard index: class                                          &                                                                  &                                                                   &                                                                    & \begin{tabular}[c]{@{}l@{}}$1.901^{***}$\\ $(0.064)$\end{tabular}   \\ 
		\hline
		Jaccard index: sub-class                                      &                                                                  &                                                                   &                                                                    & \begin{tabular}[c]{@{}l@{}}$2.497^{***}$\\ $(0.058)$\end{tabular}   \\ 
		\hline
		Jaccard index: main-group                                     &                                                                  &                                                                   &                                                                    & \begin{tabular}[c]{@{}l@{}}$3.639^{***}$\\ $(0.059)$\end{tabular}   \\ 
		\hline
		Jaccard index: sub-group                                      &                                                                  &                                                                   &                                                                    & \begin{tabular}[c]{@{}l@{}}$4.168^{***}$\\ $(0.073)$\end{tabular}   \\ 
		\hline
		&                                                                  &                                                                   &                                                                    &                                                                     \\ 
		\hline
		\rowcolor[rgb]{0.749,0.749,0.749} \textbf{Smoothing splines}  &                                                                  &                                                                   &                                                                    &                                                                     \\ 
		\hline
		&                                                                  &                                                                   &                                                                    &                                                                     \\ 
		\hline
		Publication date                                              & \multicolumn{1}{c|}{X}                                           & \multicolumn{1}{c|}{X}                                            & \multicolumn{1}{c|}{X}                                             & \multicolumn{1}{c|}{X}                                              \\ 
		\hline
		&                                                                  &                                                                   &                                                                    &                                                                     \\ 
		\hline
		Temporal difference (days)                                    & \multicolumn{1}{c|}{}                                            & \multicolumn{1}{c|}{X}                                            & \multicolumn{1}{c|}{X}                                             & \multicolumn{1}{c|}{X}                                              \\ 
		\hline
		&                                                                  &                                                                   &                                                                    &                                                                     \\ 
		\hline
		Sender citation count (log)                                   & \multicolumn{1}{c|}{}                                            & \multicolumn{1}{c|}{}                                             & \multicolumn{1}{c|}{X}                                             & \multicolumn{1}{c|}{X}                                              \\ 
		\hline
		&                                                                  &                                                                   &                                                                    &                                                                     \\ 
		\hline
		\rowcolor[rgb]{0.749,0.749,0.749} \textbf{Assestment criterions} &                                                                  &                                                                   &                                                                    &                                                                     \\ 
		\hline
		&                                                                  &                                                                   &                                                                    &                                                                     \\ 
		\hline
		AIC                                          &8229483                                         & 8202824                                          & 8138156                                          & 8058040                                            \\ 
		\hline
		GCV                                                           & 220.49                                                           & 214.69                                                            & 201.24                                                             & 185.74                                                              \\ 
		\hline
		R-squared                                           & 2.6\%                                                            & 5.17\%                                                            & 11.1\%                                                             & 18\%                                                                \\
		\hline
	\end{tabular}
\end{table}

\paragraph{Temporal effects.}
Splines associated with \emph{Model 1} indicate that the larger the temporal lag of a citation, the lower is the similarity between the citing and the cited patent. When including the temporal lag citation effect in the model specification, we are correcting for the network temporal boundaries -- i.e., for the patents issued before 1976. Related studies~\cite{kuhn_patent_2020} do not apply such a correction, with the obvious consequence that their results may be biased and do not reflect the actual changes in similarity patterns. With this correction in place, empirical results show that similarity levels start to decrease at the beginning of the 90's after experimenting an increasing trend. 

\paragraph{Citation effects}

Splines of \emph{Model 2} reveal a downward trend in the log count of senders' citations, thus suggesting that the higher the number of patent citations, the lower the pairwise similarity between citing and cited patents. The spline for the inflation of citations mitigates the effect of \emph{publication date} on patent similarity -- as the smoother decay of the corresponding spline suggests.

\emph{Model 2} shows that citations in which patents are owned by an organization tend to have lower similarity levels. This seems to suggest that organizations tend to include in their patents a series of citations that are loosely related to theirs. On the other side, citations between patents that are owned by the same organization have an important positive impact on similarity. This may be a side-effect caused by the fact that, within organizations, the office responsible for filling patent applications may be the same. As such, words and technical descriptions may be coming from the same authors who provide previous applications. Moreover, organizations involved in patent-intensive industries will likely tend to cite their own previous patents.

\paragraph{Class effects}

The importance of including technology-related information through the Jaccard similarity of IPC is highlighted by the significant increase of deviance explained and reduction of the Generalized Cross-Validation (GCV) score. Empirical estimates of \emph{Model 3} show the effects of lower levels of the patent classification scheme over the upper ones. Jaccard score computed at the IPC \emph{sub-group} level has the strongest impact on patent pairwise similarity. This should be expected given the complex structure of the IPC framework. Patents that share the lowest branch of the classification system will likely include technologies with similar features and consequently similar textual and semantic components. 

However, it is worth noting that the corresponding estimate at the \emph{section} level is higher than its \emph{class} counterpart by approximately 0.3 points. The lower error score of their respective coefficients suggests that this is not an artifact of the data, but instead, it reflects that the higher degree of the IPC has a bigger textual similarity. 

Finally, when controlling for technology similarity through Jaccard indexes, the spline associated with \emph{publication date} in \emph{Model 3} inverts its downward trend in 2011 and progressively increases up to 2021. This suggests that inside the network there could be potential interactions of patents within classes that are generating biases in the patent similarity scores. However, it is out of the scope of this paper to investigate this phenomenon. 
\paragraph{Temporal effects.} \emph{Model 1} shows that the larger the temporal lag of a citation, the lower is the similarity between the citing and the cited patent. Correcting for temporal lag to study patent similarity is important, because it corrects for the patent citation network temporal boundary -- i.e., only similarity information for patent pairs is available for patents issued after 1976. Related studies~\cite{kuhn_patent_2020} do not apply such a correction, with the obvious consequence that their results may be biased and do not reflect the actual changes in similarity patterns. With this correction in place, the results show that patent similarity levels only start to decrease at the beginning of the 1990s after experimenting an initial increasing trend. 

\paragraph{Citation effects.} \emph{Model 2} reveals a downward trend in the log count of sender citation effect. This suggests that the higher the number of patent citations, the lower the pairwise similarity between citing and cited patents. The effect due to the inflation of citations also mitigates the decline of patent similarity since the 1990s. 

\emph{Model 2} shows that citations in which patents are owned by an organization tend to have lower similarity levels. This seems to suggest that organizations tend to include in their patents a series of citations that are loosely related to theirs. On the other side, citations between patents that are owned by the same organization have an important positive impact on similarity. This may be a side-effect caused by the fact that, within organizations, the office responsible for filling patent applications may be the same. As such, words and technical descriptions may be coming from the same group of authors. Moreover, organizations involved in patent-intensive industries will likely tend to cite their own previous patents.

\paragraph{Class effects} The importance of including technology-related information through the Jaccard similarity of IPC is highlighted by the significant increase of deviance explained and reduction of the Generalized Cross-Validation (GCV) score. Empirical estimates of \emph{Model 3} show the import effect of lower levels of the patent classification scheme. Jaccard score computed at the IPC \emph{sub-group} level has the strongest impact on patent pairwise similarity. This should be expected given the complex structure of the IPC framework. Patents that share the lowest branch of the classification system will likely include technologies with similar features and consequently similar textual and semantic components. 

Finally, when controlling for technology similarity through Jaccard indices, the spline associated with \emph{publication date} in \emph{Model 3} inverts its downward trend in 2011 and progressively increases up to 2021. This is an important result, as it suggests that after accounting for important confounders, the similarity between patents is not decreasing at all, but in fact may recently be slowly increasing.

\section*{Discussion}

Patent similarity is a complex concept for which only proxy measures exist. Early approaches focused on classification-based measures, whereas more recently text-based similarity measures were introduced. In this paper, we focused our attention on a similarity measure based on SBERT, a direct evolution of the well known BERT. Nevertheless, the field of sentence embeddings is experiencing a constant development that is raising the bar in terms of model performances. A competitive alternative to SBERT could be the newly released ``Definition Sentences" (DefSent,~\cite{tsukagoshi_defsent_2021}) model, where performances of the this have proven to be marginally higher~\cite{tsukagoshi_comparison_2022}. However, at the present time, there are not enough pre-trained models on which we can provide a fair comparison. Furthermore, although training other languages model on the patent corpus~\cite{whalen_patent_2020,hain_text-embedding-based_2022} could potentially improve the accuracy of the patent similarity score, it may not change our results. In fact, although we have argued that the SBERT-based measure used in this work is a state-of-the-art approach, it is important that the main conclusions presented in this manuscript do not depend exclusively on this way of calculating patent similarity. In parallel, we have repeated our analysis with the patent similarity data used by \cite{kuhn_patent_2020}, which resulted in the same substantial results as presented in this manuscript.

When studying real-world networks, the issue of network boundaries is almost unavoidable. In the study of patent citation networks, two important boundaries we encountered in our analysis are the fact that the citing patents are US based patents from after 1976. These temporal and spatial boundaries means, first, that our concrete conclusions are firmly restricted to the modern US reality. More general conclusions would be extrapolations with more or less empirical support. It would be interesting to repeat the study in alternative jurisdictions. Secondly, particularly the temporal boundary has an important effect on the main conclusion presented in this manuscript. The sharp decline of raw patent similarity since 1976 is only reversed to an effective increase in similarity, if we accept the hypothesis that patent citation lag is an effective way to account for the temporal boundary in the patent citation network. A longer observation period would make allow us to test this assumption more carefully.

\section*{Conclusion}

Text-based similarity measures are among the most widely used indicators of patent relatedness. In this paper, we propose an efficient way to compute textual similarity scores using patent abstracts instead of entire technical descriptions. The measure we used show a similar pattern with respect to those of related studies, namely a decrease in text-based similarity starting in 1976 and continuing until recent times. The disadvantage of previous techniques is that they typically involve computationally intensive procedures that do not allow replication. In contrast, the approach of this paper avoids computational bottlenecks by making use of pre-trained NN.

Although the changes in the legal framework have had consequences on the citation process, there are other components responsible for the apparent decrease in patent similarity. Simplistic model formulations have obscured the true effect of various factors on the trend in patent similarities. Our empirical analysis focuses on the large body of patent similarities and uses a generalized linear model (GAM)~\cite{hastie_gams_1986} to uncover the non-linear relationship between several drivers on the one hand and patent similarities on the other. Explanatory variables in the model specification include both patent-related attributes and network-based measures of technological novelty. Using several model specifications, our analyses show that the observed downward trend of patent similarity scores in the last forty years is the result of several distinct endogenous effects.

The main contribution of this work concerns the analysis of possible factors that are generating the observed downward trend in patent similarity. By means of multiple GAMs, we modeled a combination of fixed and non-linear effects. What emerged from the empirical analysis is that the trend in patent similarity is affected by a series of phenomena. The most important one is the effect of time lag between citing and cited patents. This can be seen from the transition between \emph{Model 0} and \emph{Model 1}, in which we account for the time difference between the publication dates of sender and receiver. With this effect in place, the curve changes its shape and shows an interesting increase up to the mid-80s. Finally, the introduction of citation and class effects further corrects the curve by considering two other well-known phenomena: the tendency of increasing the number of citations per patent application and the increase in patent class assignments. Thanks to these adjustments, we conclude that the levels of similarity have not been constantly decreasing since 1976, but instead they show a more oscillating behavior.

%
%
%


\begin{thebibliography}{10}
	
	\bibitem{henderson1990architectural}
	Henderson RM, Clark KB.
	\newblock Architectural innovation: The reconfiguration of existing product
	technologies and the failure of established firms.
	\newblock Administrative science quarterly. 1990; p. 9--30.
	
	\bibitem{wang2019novelty}
	Wang J, Chen YJ.
	\newblock A novelty detection patent mining approach for analyzing
	technological opportunities.
	\newblock Advanced Engineering Informatics. 2019;42:100941.
	
	\bibitem{park2013identification}
	Park H, Ree JJ, Kim K.
	\newblock Identification of promising patents for technology transfers using
	TRIZ evolution trends.
	\newblock Expert systems with applications. 2013;40(2):736--743.
	
	\bibitem{wang2010identifying}
	Wang MY, Chang DS, Kao CH.
	\newblock Identifying technology trends for R\&D planning using TRIZ and text
	mining.
	\newblock R\&D Management. 2010;40(5):491--509.
	
	\bibitem{yoon2012detecting}
	Yoon J, Kim K.
	\newblock Detecting signals of new technological opportunities using semantic
	patent analysis and outlier detection.
	\newblock Scientometrics. 2012;90(2):445--461.
	
	\bibitem{verhoeven2016measuring}
	Verhoeven D, Bakker J, Veugelers R.
	\newblock Measuring technological novelty with patent-based indicators.
	\newblock Research Policy. 2016;45(3):707--723.
	
	\bibitem{veugelers2019scientific}
	Veugelers R, Wang J.
	\newblock Scientific novelty and technological impact.
	\newblock Research Policy. 2019;48(6):1362--1372.
	
	\bibitem{an2021improved}
	An X, Li J, Xu S, Chen L, Sun W.
	\newblock An improved patent similarity measurement based on entities and
	semantic relations.
	\newblock Journal of Informetrics. 2021;15(2):101135.
	
	\bibitem{kuhn_patent_2020}
	Kuhn J, Younge K, Marco A.
	\newblock Patent citations reexamined.
	\newblock The {RAND} Journal of Economics. 2020;51(1):109--132.
	\newblock doi:{10.1111/1756-2171.12307}.
	
	\bibitem{whalen_patent_2020}
	Whalen R, Lungeanu A, {DeChurch} L, Contractor N.
	\newblock Patent Similarity Data and Innovation Metrics.
	\newblock Journal of Empirical Legal Studies. 2020;17(3):615--639.
	\newblock doi:{10.1111/jels.12261}.
	
	\bibitem{an2018deriving}
	An J, Kim K, Mortara L, Lee S.
	\newblock Deriving technology intelligence from patents: Preposition-based
	semantic analysis.
	\newblock Journal of Informetrics. 2018;12(1):217--236.
	
	\bibitem{gress_properties_2010}
	Gress B.
	\newblock Properties of the {USPTO} patent citation network: 1963--2002.
	\newblock World Patent Information. 2010;32(1):3--21.
	\newblock doi:{10.1016/j.wpi.2009.05.005}.
	
	\bibitem{acemoglu_innovation_2016}
	Acemoglu D, Akcigit U, Kerr WR.
	\newblock Innovation network.
	\newblock Proceedings of the National Academy of Sciences.
	2016;113(41):11483--11488.
	\newblock doi:{10.1073/pnas.1613559113}.
	
	\bibitem{yan_measuring_2017}
	Yan B, Luo J.
	\newblock Measuring technological distance for patent mapping.
	\newblock Journal of the Association for Information Science and Technology.
	2017;68(2):423--437.
	\newblock doi:{10.1002/asi.23664}.
	
	\bibitem{younge_patent--patent_2015}
	Younge KA, Kuhn JM.
	\newblock Patent-to-Patent Similarity: A Vector Space Model.
	\newblock {SSRN} Electronic Journal. 2016;doi:{10.2139/ssrn.2709238}.
	
	\bibitem{immordino2019comparing}
	Immordino SC.
	\newblock Comparing Similarity of Patent Textual Data Through the Application
	of Machine Learning.
	\newblock University of Illinois at Chicago; 2019.
	
	\bibitem{jaffe_technological_1986}
	Jaffe A. Technological Opportunity and Spillovers of R\&D: Evidence from Firms'
	Patents, Profits and Market Value; 1986.
	\newblock Available from: \url{http://www.nber.org/papers/w1815.pdf}.
	
	\bibitem{jaffe_characterizing_1989}
	Jaffe AB.
	\newblock Characterizing the ``technological position'' of firms, with
	application to quantifying technological opportunity and research spillovers.
	\newblock Research Policy. 1989;18(2):87--97.
	\newblock doi:{10.1016/0048-7333(89)90007-3}.
	
	\bibitem{turney_frequency_2010}
	Turney PD, Pantel P.
	\newblock From Frequency to Meaning: Vector Space Models of Semantics.
	\newblock Journal of Artificial Intelligence Research. 2010;37:141--188.
	\newblock doi:{10.1613/jair.2934}.
	
	\bibitem{kuhn_information_2010}
	Kuhn JM.
	\newblock Information Overload at the U.S. Patent and Trademark Office:
	Reframing the Duty of Disclosure in Patent Law as a Search and Filter
	Problem. 2010; p.~52.
	
	\bibitem{hastie_gams_1986}
	Hastie T, Tibshirani R.
	\newblock {Generalized Additive Models}.
	\newblock Statistical Science. 1986;1(3):297 -- 310.
	\newblock doi:{10.1214/ss/1177013604}.
	
	\bibitem{deerwester_indexing_1990}
	Deerwester S, Dumais ST, Furnas GW, Landauer TK, Harshman R.
	\newblock Indexing by latent semantic analysis. 1990;41(6):391--407.
	\newblock doi:{10.1002/(SICI)1097-4571(199009)41:6<391::AID-ASI1>3.0.CO;2-9}.
	
	\bibitem{Lee_classification_2020}
	Lee JS, Hsiang J.
	\newblock Patent classification by fine-tuning BERT language model.
	\newblock World Patent Information. 2020;61:101965.
	\newblock doi:{https://doi.org/10.1016/j.wpi.2020.101965}.
	
	\bibitem{bekamiri_sbert_classification_2021}
	Bekamiri H, Hain DS, Jurowetzki R.
	\newblock Hybrid Model for Patent Classification using Augmented {SBERT} and
	{KNN}.
	\newblock CoRR. 2021;abs/2103.11933.
	
	\bibitem{hain_text-embedding-based_2022}
	Hain DS, Jurowetzki R, Buchmann T, Wolf P.
	\newblock A text-embedding-based approach to measuring patent-to-patent
	technological similarity.
	\newblock Technological Forecasting and Social Change. 2022;177:121559.
	\newblock doi:{10.1016/j.techfore.2022.121559}.
	
	\bibitem{word2vec_2013}
	Mikolov T, Chen K, Corrado Gs, Dean J.
	\newblock Efficient Estimation of Word Representations in Vector Space.
	\newblock Proceedings of Workshop at ICLR. 2013;2013.
	
	\bibitem{Liu_Cai_Guo_Chen_2021}
	Liu B, Cai Y, Guo Y, Chen X.
	\newblock TransTailor: Pruning the Pre-trained Model for Improved Transfer
	Learning.
	\newblock Proceedings of the AAAI Conference on Artificial Intelligence.
	2021;35(10):8627--8634.
	\newblock doi:{10.1609/aaai.v35i10.17046}.
	
	
	\bibitem{choi_deep_2022}
	Choi S, Lee H, Park E, Choi S.
	\newblock Deep learning for patent landscaping using transformer and graph
	embedding.
	\newblock Technological Forecasting and Social Change. 2022;175:121413.
	\newblock doi:{10.1016/j.techfore.2021.121413}.
	
	\bibitem{vaswani_attention_2017}
	Vaswani A, Shazeer N, Parmar N, Uszkoreit J, Jones L, Gomez AN, et~al.
	\newblock Attention Is All You Need.
	\newblock {arXiv}:170603762 [cs]. 2017;.
	
	\bibitem{bahdanau2014attention}
	Bahdanau D, Cho K, Bengio Y.
	\newblock Neural Machine Translation by Jointly Learning to Align and
	Translate.
	\newblock ArXiv. 2014;1409.
	
	\bibitem{rothman2021transformers}
	Rothman D.
	\newblock Transformers for Natural Language Processing: Build Innovative Deep
	Neural Network Architectures for NLP with Python, PyTorch, TensorFlow, BERT,
	RoBERTa, and More.
	\newblock Packt Publishing; 2021.
	\newblock Available from: \url{https://books.google.ch/books?id=Ua03zgEACAAJ}.
	
	\bibitem{devlin_bert_2019}
	Devlin J, Chang MW, Lee K, Toutanova K.
	\newblock {BERT}: Pre-training of Deep Bidirectional Transformers for Language
	Understanding.
	\newblock {arXiv}:181004805 [cs]. 2019;.
	
	\bibitem{reimers-2019-sentence-bert}
	Reimers N, Gurevych I.
	\newblock Sentence-BERT: Sentence Embeddings using Siamese BERT-Networks.
	\newblock In: Proceedings of the 2019 Conference on Empirical Methods in
	Natural Language Processing. Association for Computational Linguistics;
	2019.Available from: \url{https://arxiv.org/abs/1908.10084}.
	
	\bibitem{tsukagoshi_defsent_2021}
	Tsukagoshi H, Sasano R, Takeda K.
	\newblock {DefSent}: Sentence Embeddings using Definition Sentences.
	\newblock In: Proceedings of the 59th Annual Meeting of the Association for
	Computational Linguistics and the 11th International Joint Conference on
	Natural Language Processing (Volume 2: Short Papers). Association for
	Computational Linguistics; 2021. p. 411--418.
	\newblock Available from: \url{https://aclanthology.org/2021.acl-short.52}.
	
	\bibitem{tsukagoshi_comparison_2022}
	Tsukagoshi H, Sasano R, Takeda K.
	\newblock Comparison and Combination of Sentence Embeddings Derived from
	Different Supervision Signals.
	\newblock In: Proceedings of the 11th Joint Conference on Lexical and
	Computational Semantics. Association for Computational Linguistics; 2022. p.
	139--150.
	\newblock Available from: \url{https://aclanthology.org/2022.starsem-1.12}.
	
\end{thebibliography}
\end{document}